# Nonreciprocal thermal radiation based on Fibonacci quasi-periodic structures

Jun Wu[1,*], Feng Wu[2], Tiancheng Zhao[3], Han Zhai[4], and Xiaohu Wu[4,*]

1. Key Laboratory of Advanced Perception and Intelligent Control of High-end Equipment, Ministry of Education, College of Electrical Engineering, Anhui Polytechnic University, Wuhu, 241000, China

2. School of Optoelectronic Engineering, Guangdong Polytechnic Normal University, Guangzhou 510665, China

3. Beijing Aerospace Institute for Metrology and Measurement Technology, Beijing 100076, China

4. Shandong Institute of Advanced Technology, Jinan 250100, China

*Corresponding author: mailswj2011@163.com, xiaohu.wu@iat.cn

**Abstract:** To violate Kirchhoff's law is very important in the areas of thermal radiation. However, due to the weak nonreciprocity in natural materials, it is necessary to engineer novel structures to break the balance between emission and absorption. In this work, we introduce magneto-optical material into Fibonacci photonic crystals. Assisted by the nonreciprocity of the magneto-optical material and the excitation of Tamm plasmon polaritons, strong nonreciprocal thermal radiation can be realized. The difference between absorption and emission at wavelength of 16 μm can reach 0.9 at the incident angle of 60º. The distributions of the magnetic field are also calculated to verify the underlying physical origin. By engineering the parameters of the structure, it is found that strong nonreciprocal thermal radiation can be achieved at shorter wavelength and smaller incident angle. The results indicate that the Fibonacci magnetophotonic crystals are the promising candidate to engineer the nonreciprocal emission for various requirements.

# 1. Introduction

Thermal emission is a fundamental and ubiquitous process, which is related to the conversion of the thermal motion of particles in matter with non-zero temperature into electromagnetic emissions [1-4]. Breaking inversion or time reversal symmetries will have an impact on the electromagnetic waves radiated by the matter, which can be employed to control the thermal radiation and enable manipulating of radiative heat [5]. Recent discoveries in employing nonreciprocal materials have led to many novel thermal radiation properties which are quite different from that with reciprocal materials [6-9]. Besides, nanostructures made of nonreciprocal materials could violate the traditional Kirchhoff's law, i.e., breaking the balance between absorptivity and emissivity [10-24]. Nonreciprocal thermal emitters not only open up potential for manipulating fundamental properties of thermal emission, but also provide vast opportunities for applications in areas including solar energy harvesting [25]. Moreover, the generalized Kirchhoff's law, applying for both nonreciprocal and reciprocal materials, has been recently proposed and explored [26-29].

In 2014, Zhu and Fan proposed grating structures to violate the traditional Kirchhoff's law near wavelength of 16 μm around the incident angle of 60º [12]. Since the pioneer work of them, various nonreciprocal materials, mainly magneto-optical materials and Weyl semimetals, are engineered to achieve a unity difference between absorption and emission [13-22, 30]. The major of nonreciprocal systems can realize strong nonreciprocal emission at long wavelength and large angle of incidence, which greatly hinder the applications of nonreciprocal thermal radiation.

To meet different requirements, it is necessary to propose new mechanism to achieve strong nonreciprocal radiation at shorter wavelength and smaller incident angle.

To maximally violate the difference between absorption and emission, perfect absorption and emission are required. Grating structures [21], photonic crystals [31], and metamaterials [32, 33] are the promising candidates. Specifically, Fibonacci photonic crystals have attracted much attention due to its unique resonant property, which have already been applied in optical absorption [34-36], photonic band engineering [37] and omnidirectional reflection [38-40], and so on.

Here, we introduce the magneto-optical materials InAs into Fibonacci photonic crystals. Assisted by the exciting of Tamm plasmon polaritons and the nonreciprocity of the magneto-optical material, strong nonreciprocal thermal radiation can be realized, i.e., the difference between absorption and emission at wavelength of 16 μm can reach 0.9 at the incident angle of 60º. The distributions of the magnetic field are employed to verify the underlying physical origin. By engineering the parameters of the structure, it is found that strong nonreciprocal thermal radiation can be achieved at shorter wavelength and smaller incident angle. The results indicate that the Fibonacci magnetophotonic crystals are the promising candidate to engineer the nonreciprocal emission for various requirements.

## 2. Model

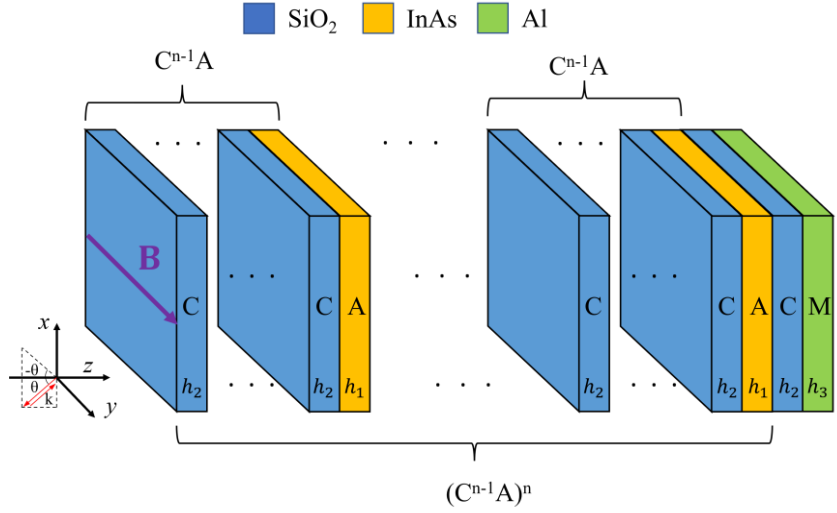

Fig. 1 Schematic of the proposed nonreciprocal thermal emitter. The magnetic field B is along the direction of y-axis.

The proposed nonreciprocal thermal emitter is shown in Fig. 1, which consists of a Fibonacci quasi-periodic magnetophotonic crystal $F_3(n)=(C^{n-1}A)^nC$ and a metal layer M, where the $n$ is the Fibonacci order [34]. The generation number of the Fibonacci magnetophotonic crystal is 3. In this work, the material of medium A is chosen to be a kind of magneto-optical materials InAs. When a magnetic field along the direction of y-axis with intensity B is imposed on the InAs, the relative dielectric constant tensor of InAs is described by [12, 17]:

$$\varepsilon = \begin{bmatrix} \varepsilon_{xx} & 0 & \varepsilon_{xz} \\ 0 & \varepsilon_{yy} & 0 \\ \varepsilon_{zx} & 0 & \varepsilon_{zz} \end{bmatrix}, \quad (1)$$

where

$$\varepsilon_{xx} = \varepsilon_{zz} = \varepsilon_\infty - \frac{\omega_p^2(\omega+i\Gamma)}{\omega\left[(\omega+i\Gamma)^2 - \omega_c^2\right]}, \quad (2)$$

$$\varepsilon_{xz} = -\varepsilon_{zx} = i\frac{\omega_p^2 \omega_c}{\omega\left[(\omega+i\Gamma)^2 - \omega_c^2\right]}, \tag{3}$$

$$\varepsilon_{yy} = \varepsilon_{\infty} - \frac{\omega_p^2}{\omega(\omega+i\Gamma)}. \tag{4}$$

Here, the detailed definitions and parameter values shown in Eqs. (2)-(4) can be found in Refs. [12] and [17]. The material of the medium C is chosen to be silica (SiO$_2$) with a refractive index of 1.45 [41]. The material of the metal M is chosen to be aluminum (Al). The relative dielectric constant of Al is obtained according to the Drude model [42]

$$\varepsilon_{Al} = \varepsilon_{\infty} - \frac{\omega_p^2}{\omega^2 + i\omega\Gamma}, \tag{5}$$

where $\varepsilon_\infty=1$, $\Gamma=1.24\times10^{14}$ rad/s and $\omega_p=2.24\times10^{16}$ rad/s [42]. The metal is optically thick, thus there is no transmission.

Since the plane of incidence is $x$-$z$ plane, no polarization conversion exists. When a plane wave with TM (transverse magnetic, with the magnetic field along the direction of $y$-axis) polarization is incident (in Fig. 1 from the left side) at an angle $\theta$, the spectral directional absorption and the spectral directional absorption emission could be obtained by [12]

$$\alpha(\theta,\lambda)=1-R(\theta,\lambda),\ e(\theta,\lambda)=1-R(-\theta,\lambda). \tag{6}$$

Here, $R(\theta,\lambda)$ and $R(-\theta,\lambda)$ are the spectral directional reflectivity at the incident angle of $\theta$ and $-\theta$, respectively. We define the difference between emission and absorption as $\eta=|\alpha-e|$, which is used to measure the magnitude of nonreciprocal radiation. During the simulation, the reflectivity is obtained by using the transfer matrix method (TMM) [43-45].

## 3. Results and discussion

The external magnetic field is fixed as 3 T in this work, which is experimentally achievable [12]. The Fibonacci order is chosen to be $n=6$. We fix the incident angle at 60º and investigate the influence of thicknesses $h_1$ and $h_2$ on the absorption and emission at wavelength of 16 μm. As illustrated in Figs. 2(a) and 2(b), the strong absorption and emission can be realized by controlling the thicknesses. The difference between emission and absorption, i.e. $\eta$, is present in Fig. 2(c). It is found that the difference is obvious at specific thicknesses.

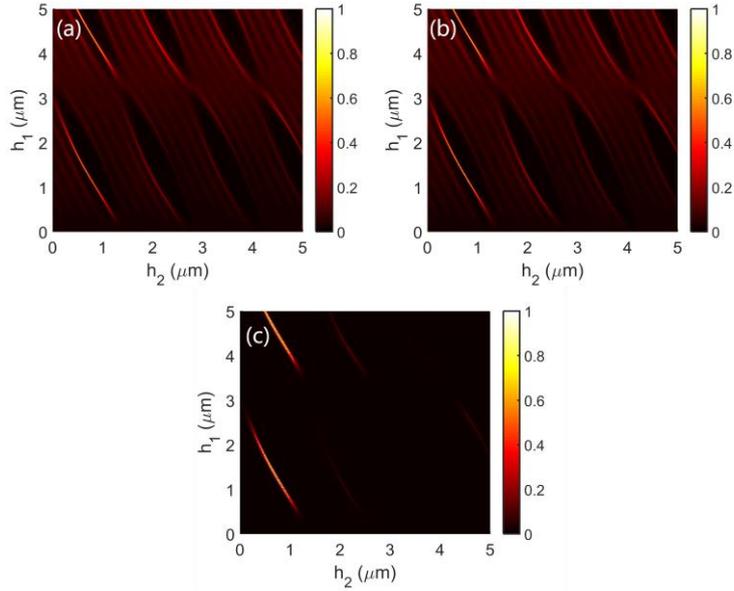

Fig. 2 The absorption (a), emission (b), and the difference between them (c) versus the changing of the thicknesses $h_1$ and $h_2$. The wavelength is 16 μm, the magnetic field is 3 T, and the incident angle is 60º.

From Fig. 2(c), we find the optimized thicknesses which can realize strong nonreciprocal thermal radiation. When $h_1$=1.5 μm, $h_2$=0.63 μm, as shown in Fig. 3(a), the absorption and emission at wavelength of 16 μm are respectively 0.95, and 0.05, resulting in near-complete violation of the traditional Kirchhoff's law. The underlying mechanism for such strong nonreciprocal thermal radiation will be disclosed later. Fig. 3(b) illustrates the absorption and emission as a function of the incident angle. It is clear the difference between absorption and emission is strong near angle of 60º. The absorption and emission are almost identical with the incident angle less than 50º.

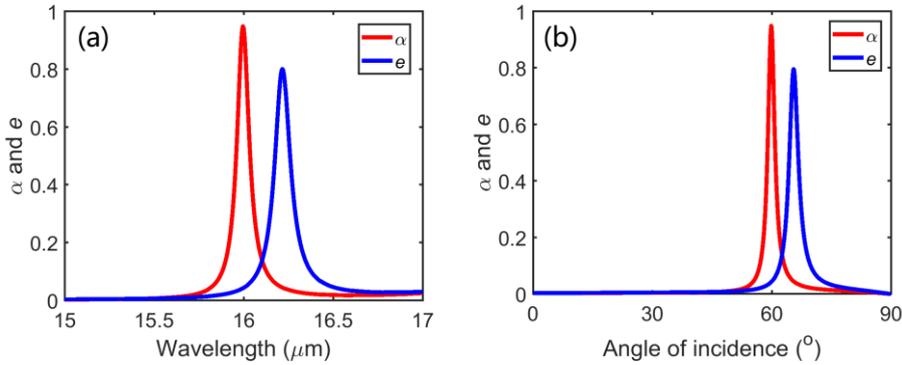

Fig. 3 (a) The absorption and emission changing with the wavelength for the incident angle of 60º. (b) The absorption and emission as a function of incident angle at the wavelength of 16 μm. The thicknesses $h_1$ and $h_2$ are 1.5 μm and 0.63 μm, respectively.

To explore the mechanism for such enhanced nonreciprocity, we calculate the magnetic field distributions ($|H_y|$) for the wavelength of 16 μm in Fig. 4. The red and blue solid lines represent the cases at $\theta=60º$ and $\theta=-60º$, respectively. Clearly, the amplitude of the magnetic field is extremely high at the interface between the Fibonacci magnetophotonic crystal and the metal layer at the incident angle of 60º, and it drops rapidly when it is away from the interface. Such phenomenon indicates the excitation of the TPPs at the interface. In the contrast, the amplitude of the magnetic field don't be enhanced at the incident angle of -60º. Although the magnetic fields are almost the same at the interface between the air and the Fibonacci magnetophotonic crystal in two cases, the reflection is totally different. For angles of incidence of 60º and -60º, the reflection coefficients are 0.13+0.20$i$ and -0.51-0.84$i$, respectively.

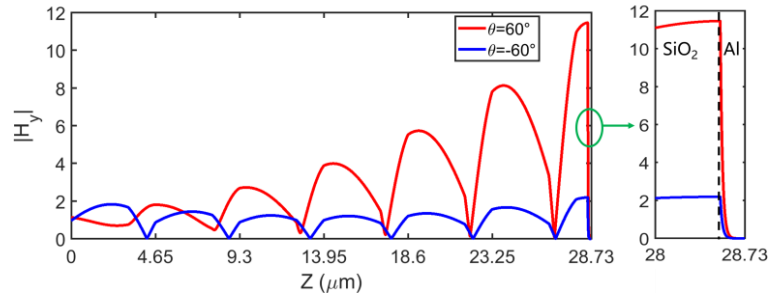

Fig. 4 The distribution of magnetic field at wavelength of 16 μm. The thicknesses $h_1$ and $h_2$ are 1.5 μm and 0.63 μm, respectively.

It is well known that the TPPs can be excited at the interface between two objects with high reflection. To further verify the excitation of the TPPs in this work, the reflection of the bare Fibonacci magnetophotonic crystal and the bare metal layer is calculated, as shown in Fig. 5. The reflection is strong at wavelength of 16 μm, either for the bare Fibonacci magnetophotonic crystal or the metal layer. Therefore, the reflection spectra could partially verify the formation of the TPPs at the interface between the Fibonacci magnetophotonic crystal and the metal layer.

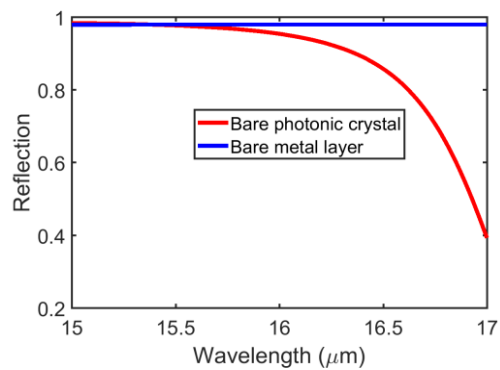

Fig. 5 The reflection of the bare photonic crystal and the metal layer. The thicknesses $h_1$ and $h_2$ are 1.5 μm and 0.63 μm, respectively.

As presented in Fig. 3(b), the nonreciprocal radiation occurs at large angle of incidence, i.e., around 60°. Here, we explore if the proposed structure can realize nonreciprocal radiation under smaller incident angle. Following the same process shown in Fig. 2, we calculated the emission and absorption as functions of the thicknesses $h_1$ and $h_2$. The wavelength is fixed at 16 μm and the angle of incidence is fixed at 30°. The optimized thicknesses can be found by looking at the difference between the absorption and emission. The results are not shown here, and we only give the optimized results. When $h_1$=1.44 μm, $h_2$=0.5 μm, the absorption and emission varying with the incident angle is shown in Fig. 6(a). It is seen that the absorption can reach 0.998, while the emission is smaller than 0.1, thus the difference between absorption and emission is also significant. The nonreciprocal radiation of the structure at angle of 10° is also explored. When $h_1$=1.23 μm, $h_2$=0.54 μm, as shown in Fig. 6(b), there is still sufficient large nonreciprocity, i.e., around 0.65.

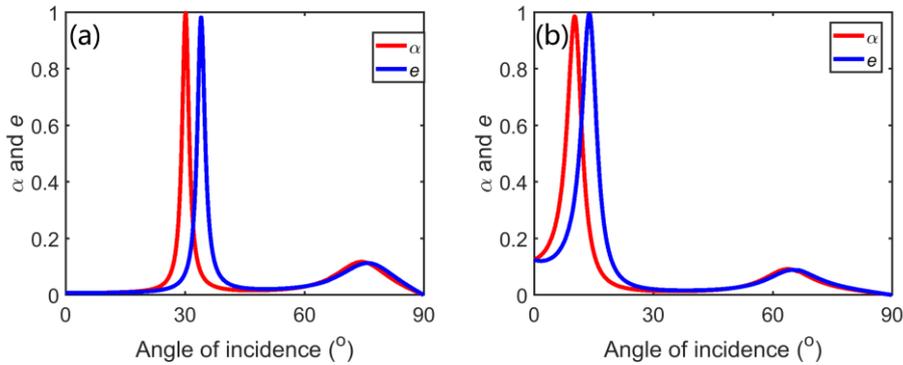

Fig. 6 The absorption and emission varying with the incident angle at wavelength of 16 μm: (a) $h_1$=1.44 μm, $h_2$=0.5 μm; (b) $h_1$=1.23 μm, $h_2$=0.54 μm.

According to the plot of the permittivity of InAs, as shown in Fig. 2 in Ref. [17], the nonreciprocal effect is enhanced as the wavelength increases. Thus, it is more challenging to achieve strong nonreciprocal thermal radiation at shorter wavelength. Following the same process shown in Fig. 2, we calculated the emission and absorption as functions of the thicknesses $h_1$ and $h_2$. The wavelength is fixed at 14 μm and the angle of incidence is fixed at 60°. When $h_1$=1.03 μm, $h_2$=0.61 μm, the absorption and emission varying with the wavelength is shown in Fig. 7(a). The difference between absorption and emission at wavelength of 14 μm can reach 0.9. When $h_1$=0.86 μm, $h_2$=0.48 μm, the absorption and emission changing with the wavelength is shown in Fig. 7(b). The difference between absorption and emission at wavelength of 12 μm can reach 0.76.

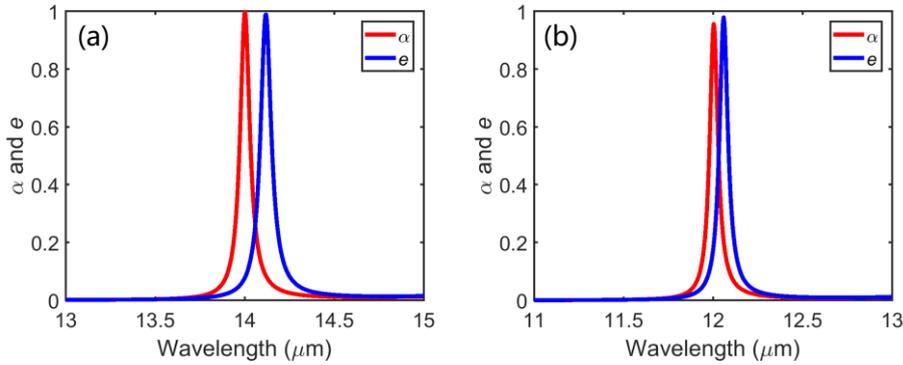

Fig. 7 The absorption and emission changing with the wavelength at the incident angle of 60°: (a) $h_1$=1.03 μm, $h_2$=0.61 μm; (b) $h_1$=0.86 μm, $h_2$=0.48 μm.

The magnetic field with B=3 T usually requires electromagnets with superconducting coils. For real applications, it should be more attractive if the magnetic field employed can be reduced [13]. When the magnetic field is 2 T, following the process in Fig. 2, we optimized the structure at wavelength of 16 μm and angle of incidence of 60°. When $h_1$=1.3 μm, $h_2$=0.72 μm, the nonreciprocity can reach 0.86, as illustrated in Fig. 8(a). When B=1 T, and $h_1$=1.32 μm, $h_2$=0.7 μm, one can see that the difference is 0.7. Therefore, the proposed structure can still obtain strong nonreciprocal radiation with B=1 T. It is worth noting that Zhao et al. designed a novel structure to break the balance between absorption and emission with a 0.3 T magnetic field. Although the magnetic field is greatly reduced, the wavelength is much larger than 16 μm [13].

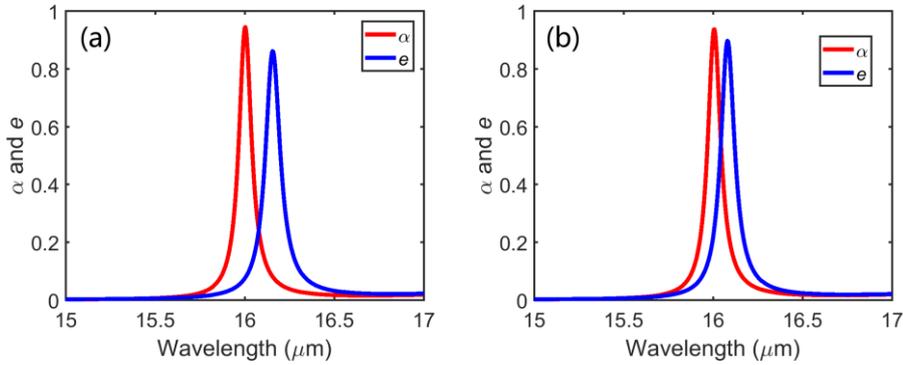

Fig. 8 The absorption and emission changing with the wavelength at incident angle of 60°: (a) $h_1$=1.3 μm, $h_2$=0.72 μm; (b) $h_1$=1.32 μm, $h_2$=0.7 μm.

## 4. Conclusions

In summary, by combining TPPs with the nonreciprocity of the magneto-optical materials, strong nonreciprocal thermal radiation is realized in Fibonacci magnetophotonic crystals. The physical origin of the strong nonreciprocal thermal radiation is revealed by the distribution of the magnetic field. The proposed structure can significantly break the balance between absorption and emission at wavelength shorter than 16 μm and small angle of incidence. It is hoped that the method proposed in this work will promote the development of novel nonreciprocal thermal emitters.

**Acknowledgements**

The authors acknowledge the support of the National Natural Science Foundation of China (Grant Nos. 61405217, 52106099 and 12104105)，the Zhejiang Provincial Natural Science Foundation (Grant No. LY20F050001), the Anhui Provincial Natural Science Foundation (Grant No. 2108085MF231), the Anhui Polytechnic University Research Startup Foundation (Grant No. 2020YQQ042), the Pre-research Project of National Natural Science Foundation of Anhui Polytechnic University (Grant No. Xjky02202003), the Natural Science Foundation of Shandong Province (Grant No. ZR2020LLZ004), and the Start-Up Funding of Guangdong Polytechnic Normal University (Grant No. 2021SDKYA033).